\documentclass[aps,prb,10pt,twocolumn,showpacs,superscriptaddress]{revtex4-2}
\usepackage{amsmath}
\usepackage{amssymb}
\usepackage{graphicx}
\usepackage{dcolumn}
\usepackage{bm}
\usepackage[utf8]{inputenc}
\usepackage{verbatim}
\usepackage{color}
\definecolor{ogreen}{rgb}{0.0, 0.4, 0.05}
\begin{document}
\title{Phase stability in SmB$_6$}

\author{M. Victoria Ale Crivillero}
\affiliation{Max-Planck-Institute for Chemical Physics of Solids,
N\"othnitzer Str. 40, 01187 Dresden, Germany}
\author{Sahana R\"{o}{\ss}ler}
\affiliation{Max-Planck-Institute for Chemical Physics of Solids,
N\"othnitzer Str. 40, 01187 Dresden, Germany}
\author{H. Borrmann}
\affiliation{Max-Planck-Institute for Chemical Physics of Solids,
N\"othnitzer Str. 40, 01187 Dresden, Germany}
\author{H. Dawczak-D\c{e}bicki}
\affiliation{Max-Planck-Institute for Chemical Physics of Solids,
N\"othnitzer Str. 40, 01187 Dresden, Germany}
\author{Priscila F. S. Rosa}
\affiliation{Los Alamos National Laboratory, Los Alamos, NM 87545, USA}
\author{Z. Fisk}
\affiliation{Department of Physics, University of California,
Irvine, CA 92697, USA}
\author{S. Wirth}
\email{Steffen.Wirth@cpfs.mpg.de}
\affiliation{Max-Planck-Institute for Chemical Physics of Solids,
N\"othnitzer Str. 40, 01187 Dresden, Germany}
\date{\today}

\begin{abstract}
We investigate flux-grown Sm-deficient Sm$_x$B$_6$ ($x < 1$) by global and
local tools, including X-ray diffraction (XRD), electronic transport, and
scanning tunneling microscopy (STM) and spectroscopy (STS). All these tools
indicate a remarkable persistence of the SmB$_6$ local structure in the
flux-grown samples even for nominal Sm concentrations as low as $x=0.75$. As
a consequence, the overall electronic properties of Sm$_x$B$_6$, and
particularly the surface conductance at low temperature, is only affected
locally by the Sm-deficiency.
\end{abstract}
\maketitle   

\section{Introduction}
Materials that host topologically non-trivial surface states have recently
become a topic of tremendous fundamental research interest \cite{has10} with
potential device applications. Usually, topological insulators can be
described by considering non-interacting electrons. However, as soon as
electronic interactions become relevant, their description quickly turns into
an often complex issue \cite{rac18}. One of the most prominent examples here
is the Kondo insulator SmB$_6$ for which topologically non-trivial surface
states have been proposed \cite{dze10,tak11}. This proposal was soon followed
by an abundance of experimental studies (see e.g.\ Refs.\
\onlinecite{li20,ros20} for a review of the vast literature). The existence
of conducting surface states is generally agreed upon by now
\cite{wol13,kim13,zha13,roe14,wol15,sye15,eo19}; however, the origin of these
surface states is less clear-cut. Besides a topological nature \cite{nxu14,
sug14,tho16,jiao18,pir20}, polarity-driven \cite{zhu13} or Rashba-split
\cite{hla18} surface states have been proposed, and issues related to bulk
in-gap states and time-dependent surface states were discussed \cite{zhu13,
he17,sen20,eo20b}. Considering the relatively simple, cubic crystal structure
(structure type CaB$_6$, $Pm\bar{3}m$), such complications came somewhat
as a surprise. Yet, there are a number of issues \cite{li20} which indeed
result in complex properties of SmB$_6$: i) the most prominent surface,
(100), is polar, ii) the Sm valence is intermediate ($\approx 2.64$ at 300 K)
and decreases slightly with decreasing temperature \cite{lut16,uts17}, iii)
a $\Gamma_8$ quartet ground state of the Sm $f^5$ configuration, is observed
experimentally \cite{sev18}, in contrast with some band structure
calculations \cite{lu13,ant02,kan15}.

Here, Scanning Tunneling Microscopy (STM) and Spectroscopy (STS) is well
versed to investigate Kondo materials due to its capability to locally explore
the surface structure and, in particular, the electronic Green's function
\cite{kir20}. Consequently, a number of STM studies have been conducted
\cite{yee13,roe14,ruan14,jiao16,miy17,jiao18,sun18,pir20,mat20,her20};
however, STM is also faced with issues mostly related to the difficulty to
cleave SmB$_6$ and the resulting scarcity of atomically flat surface areas
which, in addition, exhibit a multitude of morphologies \cite{wir20}. The
assignment of some of these surfaces is disputed, which may have consequences
for other surface sensitive measurements such as angle-resolved photoemission
spectroscopy (ARPES).

To make progress here, we study nominally Sm-deficient flux-grown samples,
Sm$_x$B$_6$ with $x \leq$ 1, in an attempt to identify and investigate 
crystallographic defects. Our results show that the SmB$_6$ structure is
formed on an atomic and mesoscopic scale with only a small number of defects,
which explains the claimed insensitivity of the bulk gap and the surface
states to off-stoichiometry \cite{eo19,li20} in flux-grown samples. Using
STS down to temperatures $T \approx 4.6$ K, we find an almost unchanged 
hybridization gap near the Fermi level, $E_{\rm F}$, for the different samples
without any sign of additional in-gap states, supporting a well-preserved
SmB$_6$ structure. Only locally, near defects, is the hybridization
diminished.

\section{Experimental}
The samples Sm$_x$B$_6$ investigated here were grown using the Al flux
technique (as detailed in Ref.~\onlinecite{rosa18}) with nominal
stoichiometries of $x =$ 0.6, 0.75, 0.9, 1.0, i.e.\ with atomic ratios of
$x:6$ in the flux. We note that the properties of Sm-deficient Sm$_x$B$_6$
grown by the floating zone method were also reported \cite{phe16,gab16,val16}.

Single-crystal x-ray diffraction (XRD) measurements were conducted on a
Rigaku AFC-7 diffraction system equipped with a Saturn 724 CCD detector using
Mo{\it K}$\alpha$ radiation ($\lambda =$ 0.71073 {\AA}) \cite{xrd}. Resistance
measurements were performed using a Physical Properties Measurement System
(PPMS) by Quantum Design, Inc.

STM/STS was conducted in an ultra-high vacuum system \cite{omi} at pressures
$p\lesssim 2.5 \!\times \! 10^{-9}$ Pa and at temperatures $T \gtrsim 4.6$
K (if not stated otherwise, the presented STM/STS results were acquired at
base temperature). A total of 9 samples were cleaved {\it in situ} at
temperatures $T \sim 20$ K approximately along one of the principal cubic
crystallographic axes; we here report results on 4 samples with $x = 0.75$,
0.9. On the remaining samples, atomically flat surface areas could not be
found. The tunneling current $I$ was measured using tungsten tips and a bias
voltage $V_{\rm b}$ was applied to the sample. Most topographies were obtained
in dual bias mode, i.e.\ forward and backward scan along the fast scan
direction were obtained with different $V_{\rm b}$. The ${\rm d}I/{\rm d}V$
spectra were acquired by a lock-in technique applying a modulation voltage of
typically $V_{\rm mod} =$ 0.3 mV at 117 Hz (exceptions are noted in the
respective figure caption).

We emphasize that, whenever possible, {\em identical} samples were used for
the different measurements.

\section{Results}
\subsection{X-ray diffraction}
In an effort to complement our local STM/S measurements, x-ray diffraction
(XRD) was performed on some samples which were used later for STM/S. The
\begin{table}[t]
\caption{Nominal composition and refined Sm occupancy, lattice constant $a$,
refinement parameters, and atomic distances of the Sm-deficient samples
Sm$_x$B$_6$ at 300~K.}
\label{tab1}
\begin{ruledtabular}
\begin{tabular}{llll}
sample / batch &  \#1           &  \#2             &  \#3           \\
\hline
\rule{0pt}{10pt}
nominal composit. & Sm$_{0.9}$B$_6$ & Sm$_{0.75}$B$_6$ & Sm$_{0.6}$B$_6$\\
\hline
\rule{0pt}{10pt}
refined Sm occup.  & 0.983(10)  & 0.990(14)        & 0.968(17)      \\
$a$ (\AA)     & 4.1393(3)       & 4.1387(3)        &  4.1385(2)     \\
\# of unique  &                 &                  &                \\
reflections   &  82             &  81              &  82            \\
\# of refined &                 &                  &                \\
parameters    &  7              &  7               &  7             \\
$R$           &  0.0069         &  0.0105          &  0.0110        \\
\hline
\rule{0pt}{10pt}distances  &    &                  &                \\
Sm--B (\AA)   &  3.0413(4)      &  3.0408(6)       &  3.0412(7)     \\
B--B (\AA) int$^a$ & 1.7584(18) &  1.758(3)        &  1.756(3)      \\
B--B (\AA) ext$^b$ &  1.653(3)  &  1.652(4)        &  1.655(5)
\footnotetext{intra-octahedral B-distances}
\footnotetext{inter-octahedral B-distances}
\end{tabular}
\end{ruledtabular}
\end{table}
results, as summarized in Table \ref{tab1}, indicate very good
consistency and reflect particular efforts to allow for refinement of the site
occupancy of Sm atoms as part of the structural model. This implies that there
are no significant differences among the domains forming the crystallites
under investigation. More importantly, however, these crystallites exhibit
very little deviation from perfect stoichiometry. Even for the nominally most
Sm-deficient sample \#3, the refined Sm reaches about 97\% occupancy, i.e.,
the {\it intrinsic} deficiency is only about 3\%. Consequently, boron-rich
phases are either left behind in the Al-flux during the growth or accumulate
between crystallites within the sample, or a combination of both. Also, there
appears to be no direct correlation between the nominal and the refined
composition as the refined composition of sample \#2 (nominally $x =$ 0.75) is
found to be closer to the ideal stoichiometry than sample \#1 (nominally $x =$
0.9). We take this as an indication that {\em flux-grown} SmB$_6$ is
thermodynamically very stable and therefore tends to be stoichiometric. This
conjecture is in line with earlier reports on flux-grown and floating-zone
grown SmB$_6$ which found a comparatively larger amount of Sm-deficiency in
the latter samples \cite{phe16,ghe19}, specifically if the samples were
remelted \cite{ore17}.

Given the here established large discrepancy and the apparent absence of
correlation between nominal and refined compositions of our Sm-deficient
samples, we will just refer to sample (or batch) number as
provided in Tab.\ \ref{tab1} for the remainder of this report.

\subsection{Transport measurements}
\label{sec-res}
\begin{figure}[t]
\hspace*{-0.3cm}\includegraphics[width=0.4\textwidth]{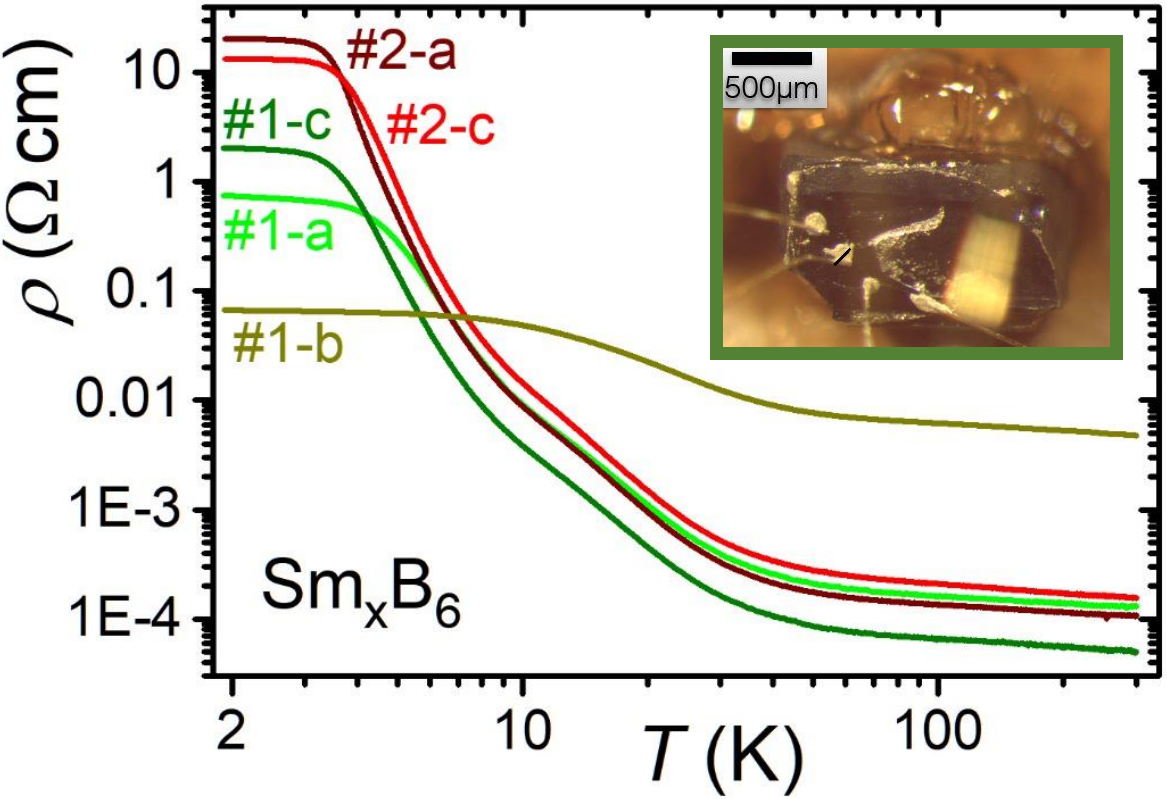}
\caption{Resistivities $\rho$ of samples \#1 and \#2; the letters specify:
a - as grown, b - bent, c - cleaved. The much smaller change of $\rho$ in
curve \#1-b is likely due to a bending of sample \#1 during cleaving (see
text). Inset: Photograph of cleaved sample \#1. The cleaved part (left) of
the surface appears shiny and contains the contacts.}
\label{resist}  \end{figure}

The resistivities of one exemplary sample \#1 and one sample \#2, which
were also used in our XRD and STM investigations, are summarized in Fig.\
\ref{resist}. In general, $\rho (T)$ follows the overall behavior found for
pristine SmB$_6$ and is in good agreement with previously published data,
in particular \cite{sta19} and therefore, shall not be discussed in detail
here. Rather, we focus on the impact of the cleaving procedure (as applied
for STM measurements) on the sample properties. To this end, Fig.\
\ref{resist} compares $\rho (T)$ of samples \#1 and \#2 before {denominated
\mbox{``-a''}}, as-grown) and after cleaving and STM investigation
{denominated ``-c'', cleaved, see inset for cleaved surface of sample \#1 with
contacts attached). Clearly, for sample \#2 there is no significant change.
For sample \#1 (with refined stoichiometry further away from 1:6 compared to
sample \#2) an increase of $\rho (T)$ over more than four orders of magnitude
upon cooling is only observed after cleaving, curve \#1-c in Fig.\
\ref{resist}. We speculate that the off-stoichiometry of our samples may
influence the surface (and contact) quality in the as-grown case, curve \#1-a.

In one case, a significantly smaller increase of resistivity was found, curve
\#1-b. Closer microscopic inspection revealed a bending of the cleaved, tiny
sample after dismounting from our STM sample holder and mounting for transport
measurements. Such bending may result in a strained sample. For SmB$_6$ under
strain, a considerably reduced increase of resistance with cooling and a
higher temperature below which the surface state dominates electrical
transport was reported \cite{ste17}. Indeed, the approach to a low-$T$
saturation of $\rho(T)$ appears at higher temperature, around 8~K, and
$\rho(T)$ is significantly enlarged at room temperature.

\subsection{STM on sample \#1}
As was reported before \cite{roe16,sun18,wir20,mat20,her20} large atomically
resolved surface areas are rarely observed on pristine SmB$_6$ and typically
have to be searched for extendedly. Finding a flat, clean area appears to be
slightly easier in case of Sm$_x$B$_6$ with nominal $x<$ 1 indicating that
some defects are present in our Sm-deficient samples and promote cleavage.
This observation is in line with a reported decrease of hardness of
Sm$_x$B$_6$ as $x$ decreases \cite{eo19} and the recent suggestion
\cite{eo20b} that these defects are line defects.

An atomically resolved topography over an area of $(15 \times 11)$ nm$^2$ of
sample \#1 is presented in Fig.~\ref{topo09}(a). Despite some defects, two
flat areas can be distinguished which are separated by a step edge. Its height
of about 0.21 nm can be inferred from the pink line scan in
Fig.~\ref{topo09}(b) taken along the line of similar color in (a). The
distances between corrugations conforms to the lattice constant $a$ and hence,
$(1\times 1)$ terminations are observed within both flat areas I and II
corresponding to either Sm or B$_{(1)}$/B$_{(6)}$ surfaces. At first glance,
the observed step height of about 0.21 nm $\approx a/2$ conforms well with the
expected height difference between Sm and B$_{(2)-(5)}$ terminations (see
inset of Fig.~\ref{topo09}(b) for B notations). However, such an assignment
would involve different terminations (Sm vs.\ broken B octahedra) and hence,
would call for different appearances and arrangements of the corrugations on
surfaces I and II (including the so-called doughnuts \cite{ruan14}). In
contrast, we observe similar heights and distances of the corrugations on both
terraces. Assuming the flat surface areas I and II coincide with B$_{(6)}$ and
B$_{(1)}$ terminations, respectively, a step height of 0.248 nm is expected.
This value exceeds the measured step height by almost 20\%. We note that
Fig.\ \ref{topo09}(a) was obtained in dual bias mode, $V_{\rm b} = \pm 35$ mV
without noticeable difference between these $V_{\rm b}$-values.

To scrutinize the step further, the height maxima (marked by $|$) within area
I and II along the pink line were analyzed separately. Assuming constant
distances $d$ within each one of the terraces yields $d_{\rm I} = 0.427$ nm
and $d_{\rm II} = 0.430$ nm, deviating less than 4\% from $a$. Using these
$d$-values, the expected atomic positions can be interpolated (marked by
$\bullet$) which deviate in lateral position from the measured height maxima
by less then 1\% of $a$. This accuracy allows for an extrapolation of the
\begin{figure}[t]
\includegraphics*[width=0.42\textwidth,viewport=-60 0 376 276]{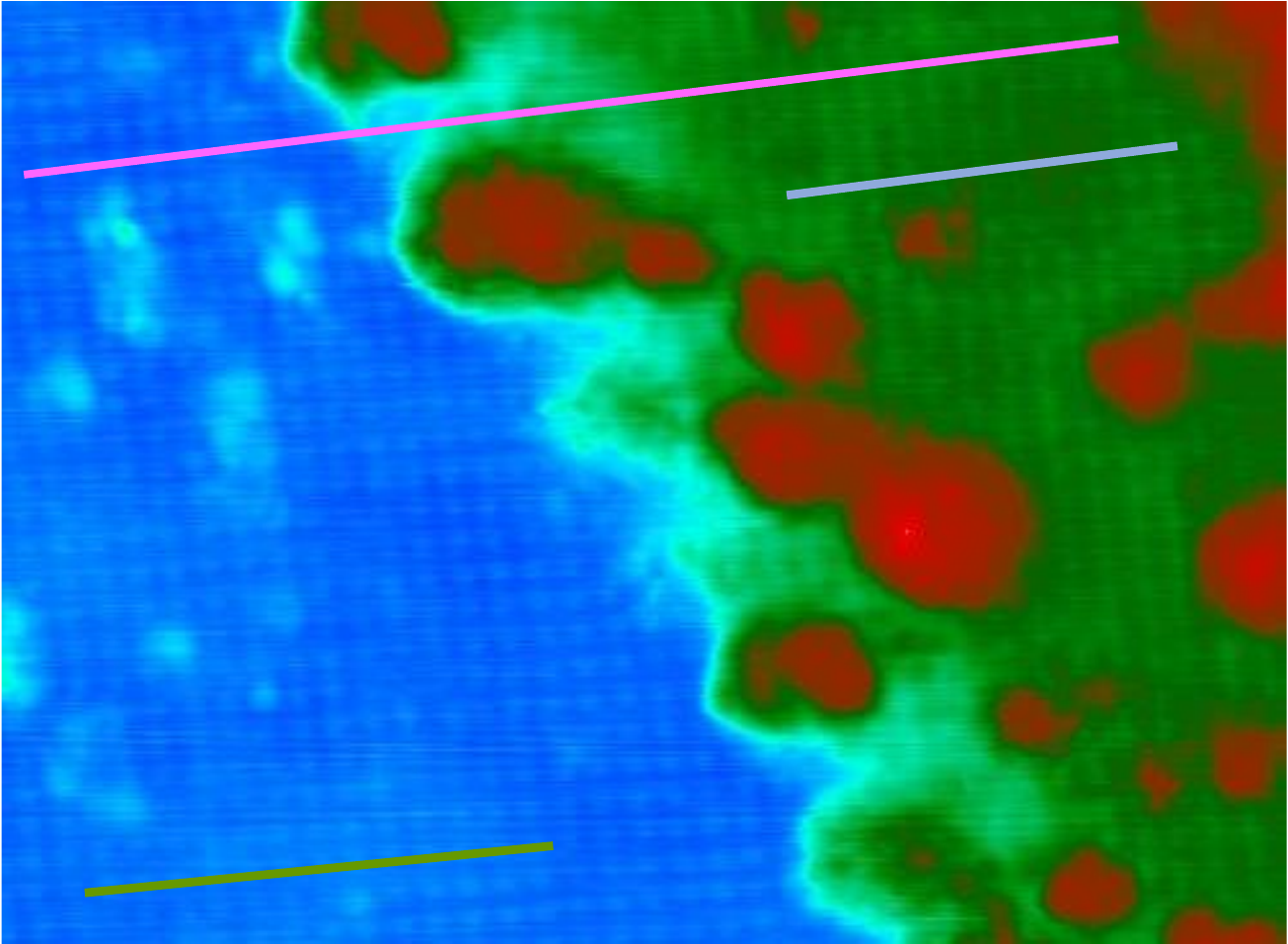}
\includegraphics*[width=0.48\textwidth]{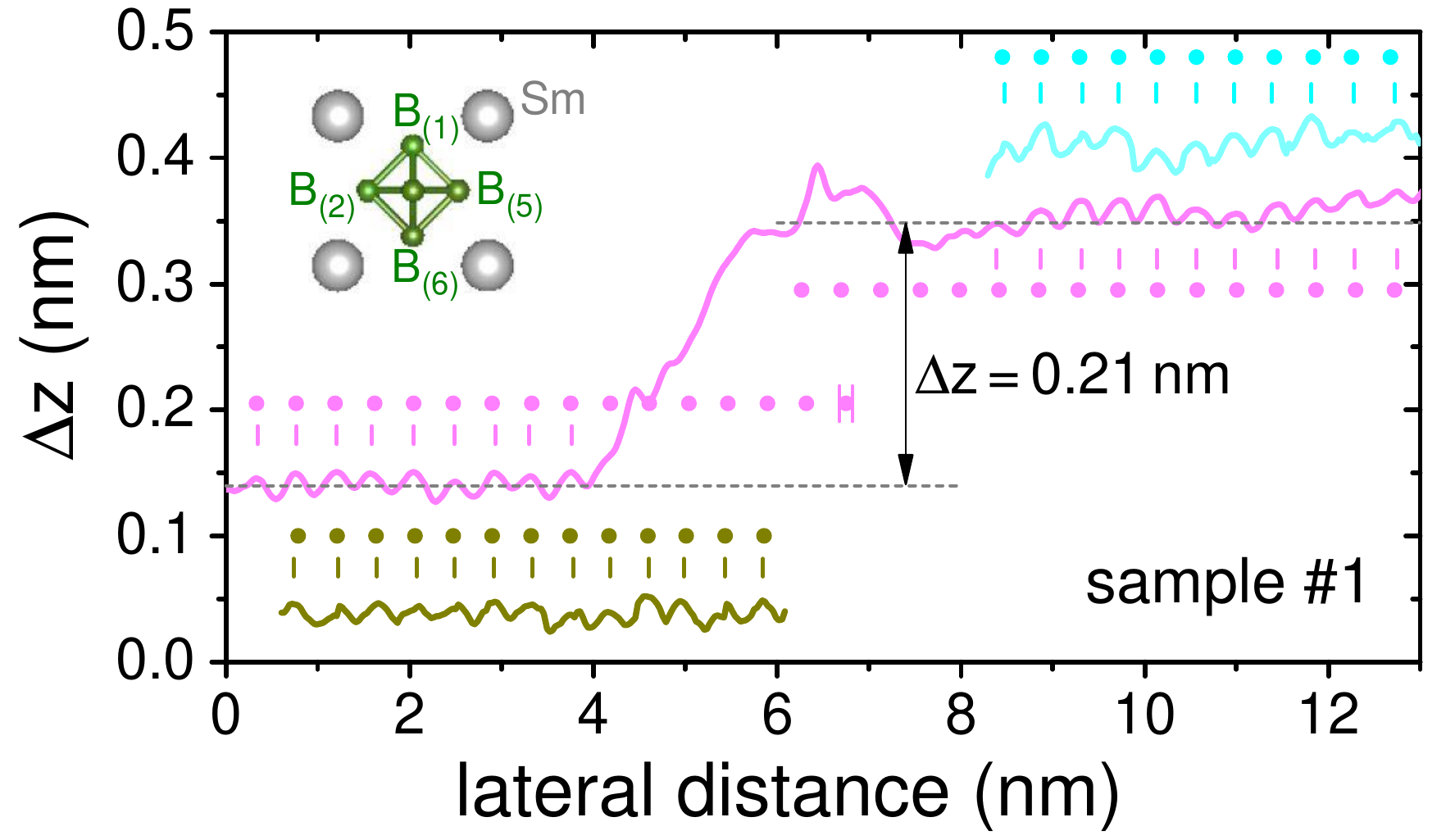}
\unitlength1cm \begin{picture}(-0.4,2)
\put(-8.1,9.4){\sffamily\bfseries\large (a)}
\put(-8.7,4.4){\sffamily\bfseries\large (b)}
\put(-7,7.2){\sffamily\bfseries \textcolor{white}{I}}
\put(-1.7,7.2){\sffamily\bfseries \textcolor{white}{II}}
\put(-7.2,2.8){I}
\put(-0.8,2.7){II}
\end{picture}
\caption{(a) STM topography (15 nm$\,\times\,$11 nm) of two atomically
resolved areas (I and II) on sample \#1; the separating step is possibly
related to a line defect. Corrugations within the flat areas are spaced by the
lattice constant $a$. $V_{\rm b} = 35$~mV, $I_{\rm sp} = 0.15$ nA; (b) Height
scans along lines of similar color indicated in (a). Distances between
corrugations (marked by $|$) were interpolated to estimate atomic positions
($\bullet$). The step height (pink line) is $\sim$0.21 nm.} \label{topo09}
\end{figure}
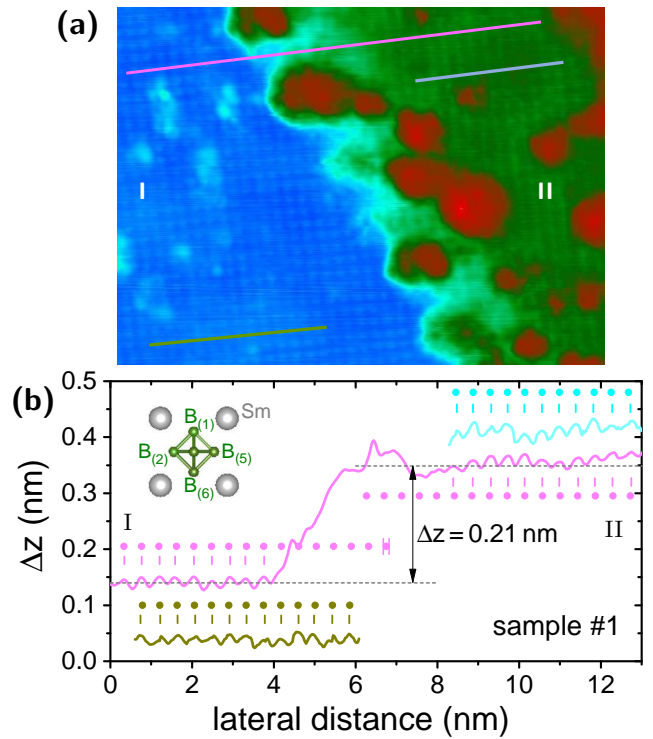
expected atomic positions into the region of the step, i.e.\ beyond the
observable height maxima. In order to estimate the error in our extrapolation,
additional line scans on both areas were evaluated (green and light blue line
scans) yielding $d_{\rm I}^{\rm green} = 0.422$ nm and $d_{\rm II}^{\rm blue}
= 0.423$ nm. This spread of the $d$-values is included in the error bar
($\pm 0.07$ nm) of the extrapolated atomic positions, see the right-most pink
marker of surface I. Within the error of this extrapolation the atomic
positions overlap without offset suggesting a certain crystalline continuity
in the present field of view.  All this affirms that the clean surface areas
represent topographies as expected from largely undisturbed crystalline
SmB$_6$ without indication for considerable Sm deficiency. The latter is in
accord with our findings from XRD.

To gain further insight, zoomed-in areas on either side of the step seen in
Fig.\ \ref{topo09}(a) are presented in Fig.\ \ref{workfun}(a)--(d). These
images were acquired in dual-bias mode, $V_{\rm b} = \pm 0.035$~V, allowing
for direct comparison of the different $V_{\rm b}$-values. As shown for two
line scans in Fig.\ \ref{workfun}(f), the apparent height difference due to
the different $V_{\rm b}$ is less than 10 pm. While the undisturbed lattice
regions appear little influenced by $V_{\rm b}$, the defects are slightly more
pronounced for negative $V_{\rm b} = -0.035$~V indicating more negatively
charged defects with respect to their surroundings. This, together with the
small total height of the defect, rules out a Sm adatom generating this
defect. The defect may be caused by either lattice imperfections within or
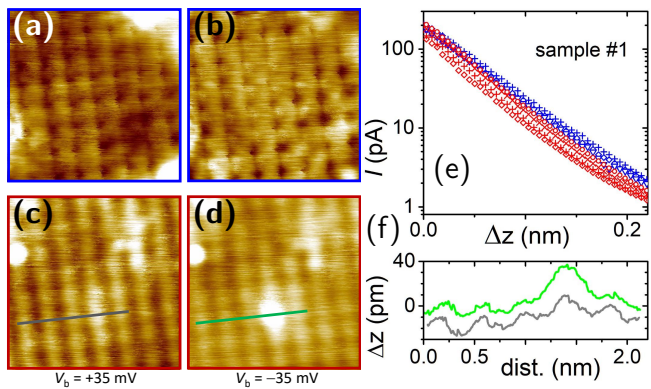
\begin{figure}[t]
\hspace*{-0.3cm}\includegraphics[width=0.48\textwidth]{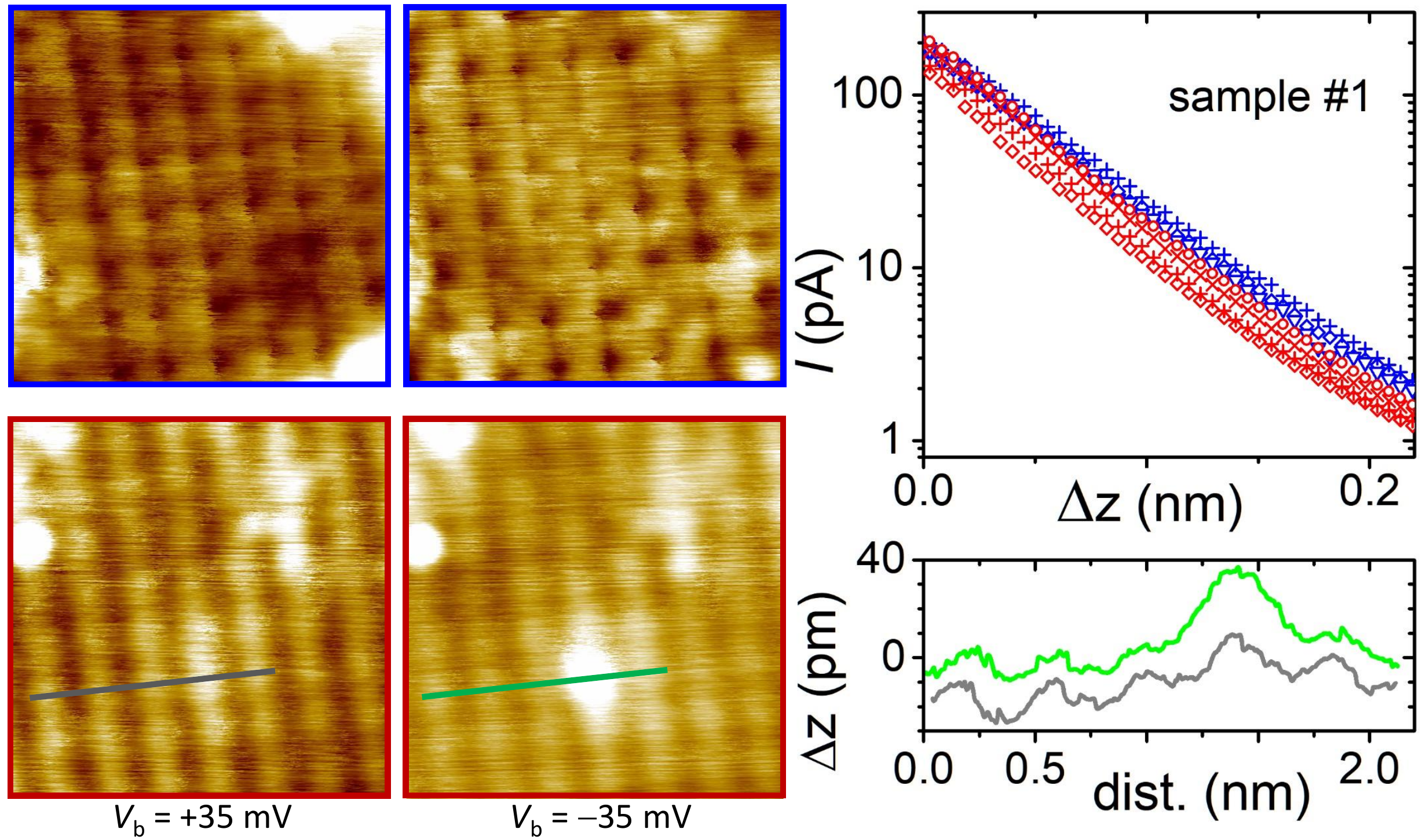}
\unitlength1cm \begin{picture}(-0.4,2)
\put(-8.6,4.7){\sffamily\bfseries\large \textcolor{white}{(a)}}
\put(-6.2,4.7){\sffamily\bfseries\large (b)}
\put(-8.6,2.2){\sffamily\bfseries\large (c)}
\put(-6.2,2.2){\sffamily\bfseries\large (d)}
\put(-3.0,2.8){\sffamily\large (e)}
\put(-3.9,2.0){\sffamily\large (f)}
\end{picture}
\caption{(a)-(d) Topographies (3 nm$\,\times\,$3 nm) on the upper side II
[(a), (b), blue frame] and the lower side I [(c), (d), red frame] of the
same terrace, but outside the field-of-view of Fig.\ \ref{topo09}(a). Images
taken quasi-simultaneously in dual-bias mode: (a), (c) $V_{\rm b} = +35$~mV;
(b), (d) $V_{\rm b} = -35$~mV. (e) Representative current $I$ vs. tip
displacement $\Delta z$ curves taken within the areas of corresponding colors.
(f) Height profiles along the lines of corresponding color indicated in (c)
and (d).} \label{workfun}  \end{figure}
below the surface, or B atoms/cluster resulting from the cleaving process. We
note that there is no contrast reversal observed for the $V_{\rm b}$-values
used here, neither within the lower area I nor the upper area II.

Surfaces are characterized by their work functions $\Phi_{\rm s}$. A related
parameter, the tunneling barrier height $\Phi$, can be studied by measuring
the tunneling current $I$ in dependence on tip-sample distance $\Delta z$. In
clean cases, $\Phi$ can be estimated from $I(z) \propto\exp (-2 \kappa \,
\Delta z)$ with $\kappa^2 = \frac{2 m_e}{\hbar^2} \Phi$, where $m_e$ is the
bare electron mass and $V_{\rm b} \ll \Phi_{\rm s,t}$. Here, $\Phi_{\rm t}$
is the tip work function. A few $I$ vs. $\Delta z$-curves are presented in
Fig.\ \ref{workfun}(e) taken on numerous defect-free spots on both surfaces.
The barrier heights from the lower surface I (red markers) range between 5.8
eV $\leq \Phi_{\rm I} \leq$ 6.7 eV, while on the upper surface (blue markers)
4.5 eV $\leq \Phi_{\rm II} \leq$ 5.4 eV. Albeit there appears to be a
difference in the barrier heights within the two surface areas, the total
range of $\Phi$ is remarkably close to the one obtained on B-terminated
EuB$_6$ \cite{wir20}. Moreover, our range of $\Phi$ is considerably smaller
than the one reported in \cite{sun18}, possibly due to the much larger clean
areas investigated. Specifically, the local barriers heights for clean areas
were reported to be of order 4 eV \cite{sun18}, while our observations indicate
somewhat larger values. According to Ref.\ \onlinecite{sun18}, a small work
function of 2 eV is expected for Sm-terminated surfaces, a value close to 2.7
eV for pure Sm \cite{mic77}, whereas on B-terminated surfaces it should be at
least twice as high. The latter is in line with pure B (4.45 eV, \cite{mic77})
and an early report on SmB$_6$ with very little Sm in the surface layer
(4.2 eV, \cite{aon79}) as well as more recent angle-resolved photoemission
\begin{figure}[t]
\hspace*{-0.3cm}\includegraphics[width=0.40\textwidth]{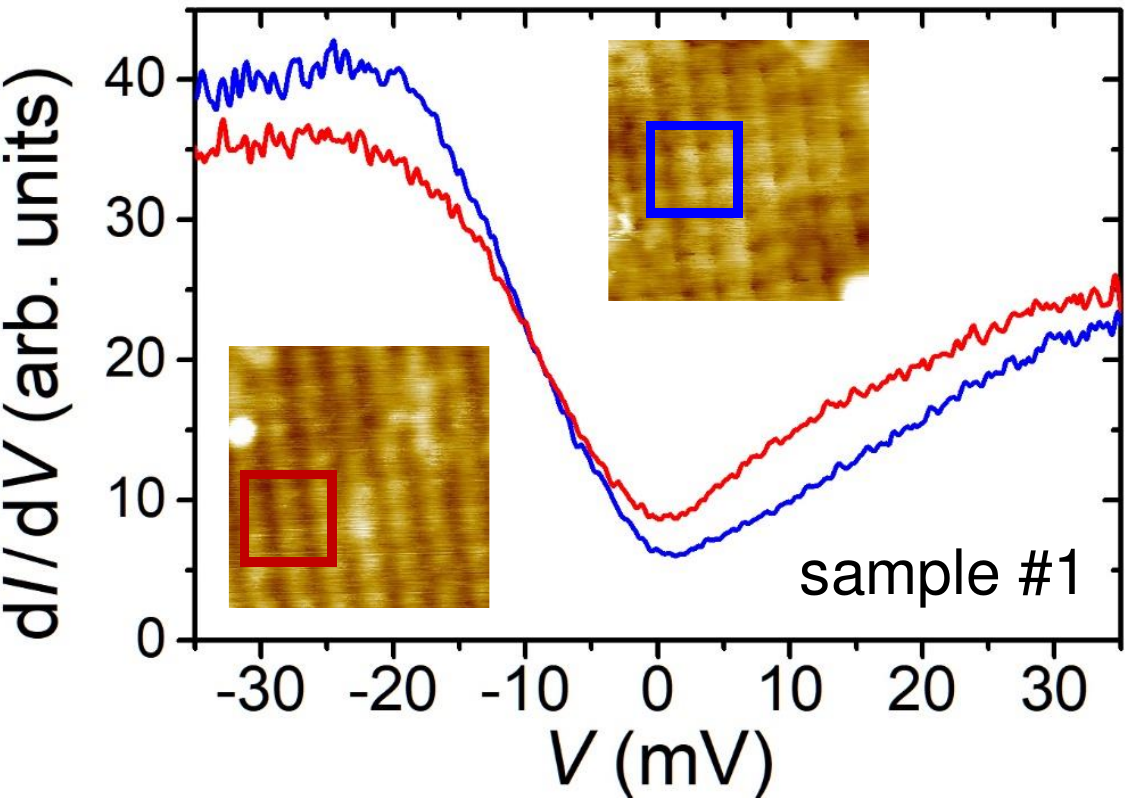}
\caption{Tunneling spectra obtained on surfaces I and II, respectively.
Insets: Topographies with areas marked where the spectra were taken [same
parameters as in Fig.\ \ref{workfun}(a), (c)]. Spectra were obtained on a
$4\times 4$ grid and averaged ($V_{\rm b} = +0.2$~V, $I_{\rm sp}
=$ 0.15 nA, $V_{\rm mod} = 0.15$~mV).}
\label{specI+II}  \end{figure}
measurements (4.5 eV, \cite{neu13}). Although we can only measure $\Phi$,
rather than $\Phi_{\rm s}$, these results support an assignment of both
surfaces I and II to B-terminations, in line with our earlier results \cite{roe14,roe16,wir20}.

To further support this assignment, STS was conducted within both areas. As
evidenced by Fig.\ \ref{specI+II}, there is little difference between the
spectra on both areas. While the prominent peak at around $-7$ mV observed on
clean areas and at lower $T$ \cite{jiao16,sun18} is absent, they exhibit the
reduced local density of states near $V= 0$ typical of the Kondo hybridization
in SmB$_6$. In addition, a broad, yet moderate, hump at around $-20$ mV is
observed \cite{miy17,sun18,her20}. The Kondo hybridization \cite{wir16} in
SmB$_6$ allows for co-tunneling into Sm 4$f$ states and the conduction band
that can give rise to a much more pronounced peak at this energy \cite{yee13,
ruan14,roe14,jiao16,sun18,pir20,her20}. The small hump then indicates very
little tunneling into the Sm 4$f$ states \cite{roe14,sun18}. This reinforces
our finding above that both surfaces I and II are likely B terminated. In
addition, it was suggested that a maximum in ${\rm d}I/{\rm d}V$ at $-20$ mV
may result from a local doping effect due to boron clusters on the surface
\cite{sun18} which is in line with our assignment of the surface defects to
B. Note also that the Sm-deficiency of this sample \#1 is likely supportive
in establishing these surface properties.

Figure \ref{topo42}(a) presents another clean surface area which exhibits some
defects similar to those reported as doughnuts \cite{ruan14}. However, their
origin appears to be different from \cite{ruan14} for two reasons: i) the
height scans along a $\langle$100$\rangle$ direction, Fig.\ \ref{topo42}(c),
reveal two protrusions almost a lattice constant apart. ii) individual
doughnuts are observed exclusively. The center of the doughnuts are located
on top of dents of the underlying lattice. This positioning, along with the
central dent of the doughnuts [clearly seen in Fig.\ \ref{topo42}(b)],
suggests that the defects are {\em not} caused by single Sm-atoms on top of a
B-termination or vice versa. Rather, they are likely made up of several
(conceivably four) B$_{(6)}$ atoms [cf.\ side view Fig.\ \ref{topo42}(d)]
\begin{figure}[t]
\hspace*{-0.3cm}\includegraphics[width=0.44\textwidth]{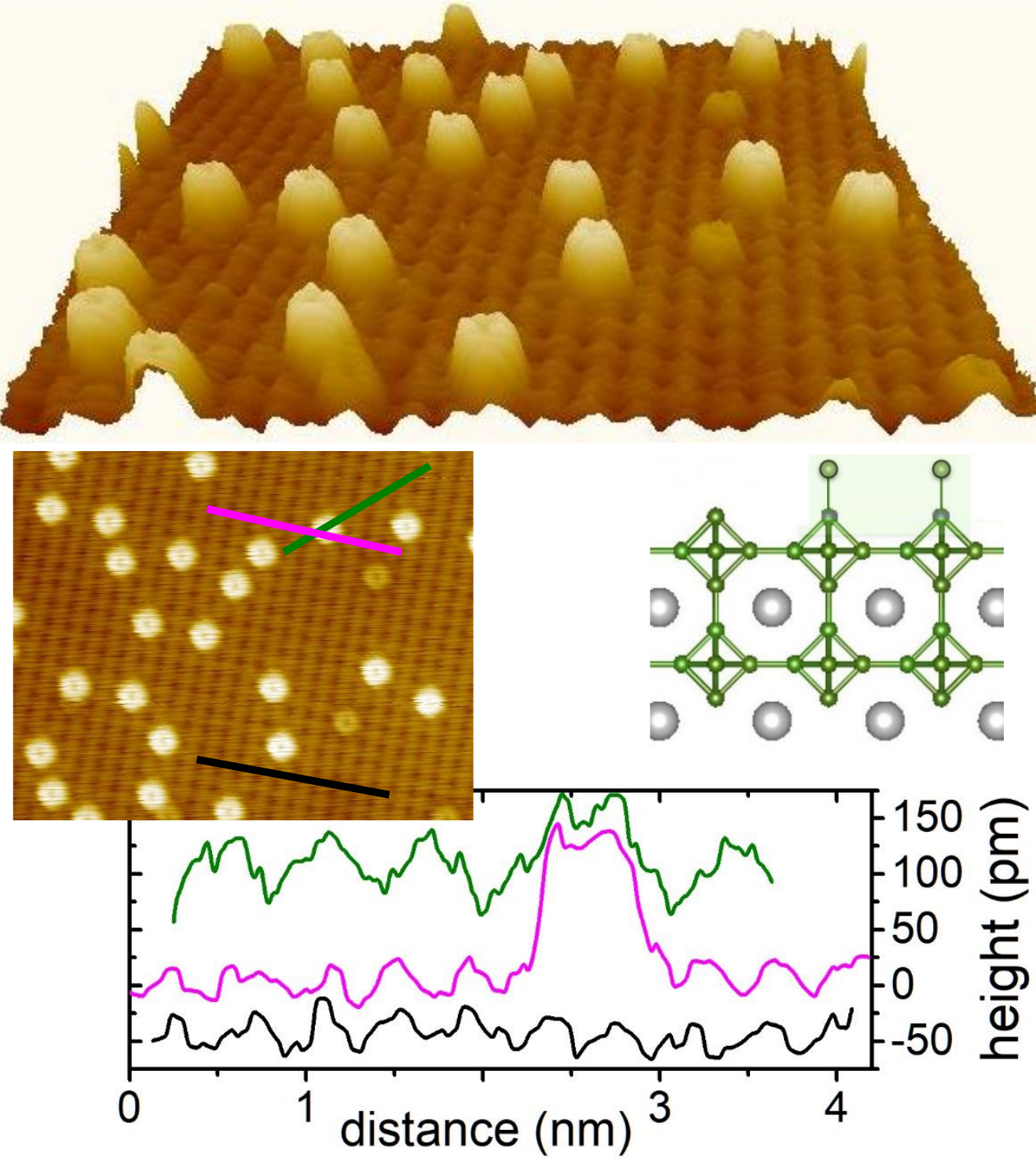}
\unitlength1cm \begin{picture}(-0.4,2)
\put(-7.6,8.2){\sffamily\bfseries\large (a)}
\put(-7.95,5.05){\sffamily\bfseries\large \textcolor{white}{(b)}}
\put(-3.8,4.8){\sffamily\bfseries\large (d)}
\put(-7.7,1.9){\sffamily\bfseries\large (c)}
\put(-1.55,5.1){\sffamily \textcolor{ogreen}{B$_{(6)}$}}
\end{picture}
\caption{(a) Topography (10 nm$\,\times\,$8 nm) on sample \#1 with
few defects on an otherwise clean (1$\times$1) surface. $V_{\rm b} = +0.2$~V,
$I_{\rm sp} =$ 0.5 nA. (b) Same area with the positions of the height scans
presented in (c) indicated. (d) Side view of a possible doughnut formation
(see also Fig.\ \ref{topo09}(b) for B$_{(6)}$ assignment).}
\label{topo42}  \end{figure}
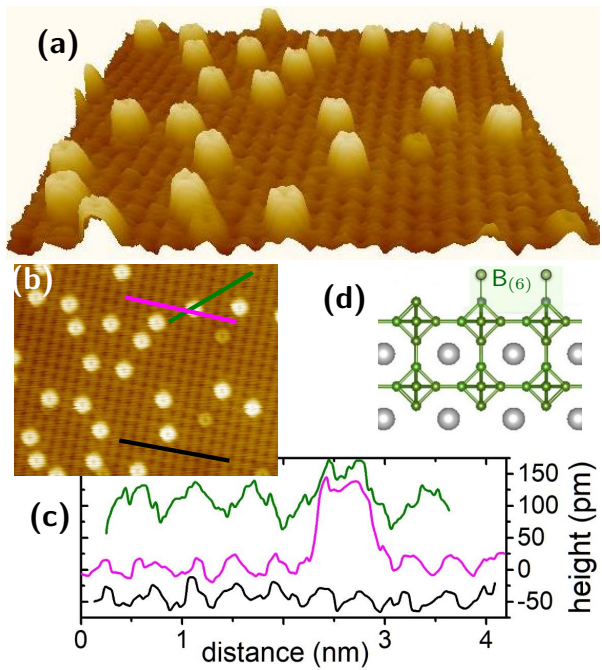
forming a round structure, as nicely seen in the 3-dimensional topography Fig.\
\ref{topo42}(a). This is supported by their apparent height [about 130 pm,
magenta line scan in Fig.\ \ref{topo42}(c)] which is near the inter-octahetral
B-distance. Moreover, the size of the doughnuts, both along the
$\langle$100$\rangle$ and the $\langle$110$\rangle$ direction, render a
pentaboride cluster on the surface unlikely. Note that the off-stoichiometry
of sample \#1 makes excess B on the surface likely. Nonetheless, it remains
unclear why an apparently invariable number of B-atoms may form such doughnut
structures. We speculate that this results from the energetically high impact
of the cleaving process. It is worth noting that there are similarities to the
topographies presented in \cite{ruan14,sun18} albeit with a much higher defect
density there.

A revealing type of defect is presented in Fig.\ \ref{cross}. The most common
\begin{figure}[t]
\hspace*{-0.3cm}\includegraphics[width=0.48\textwidth]{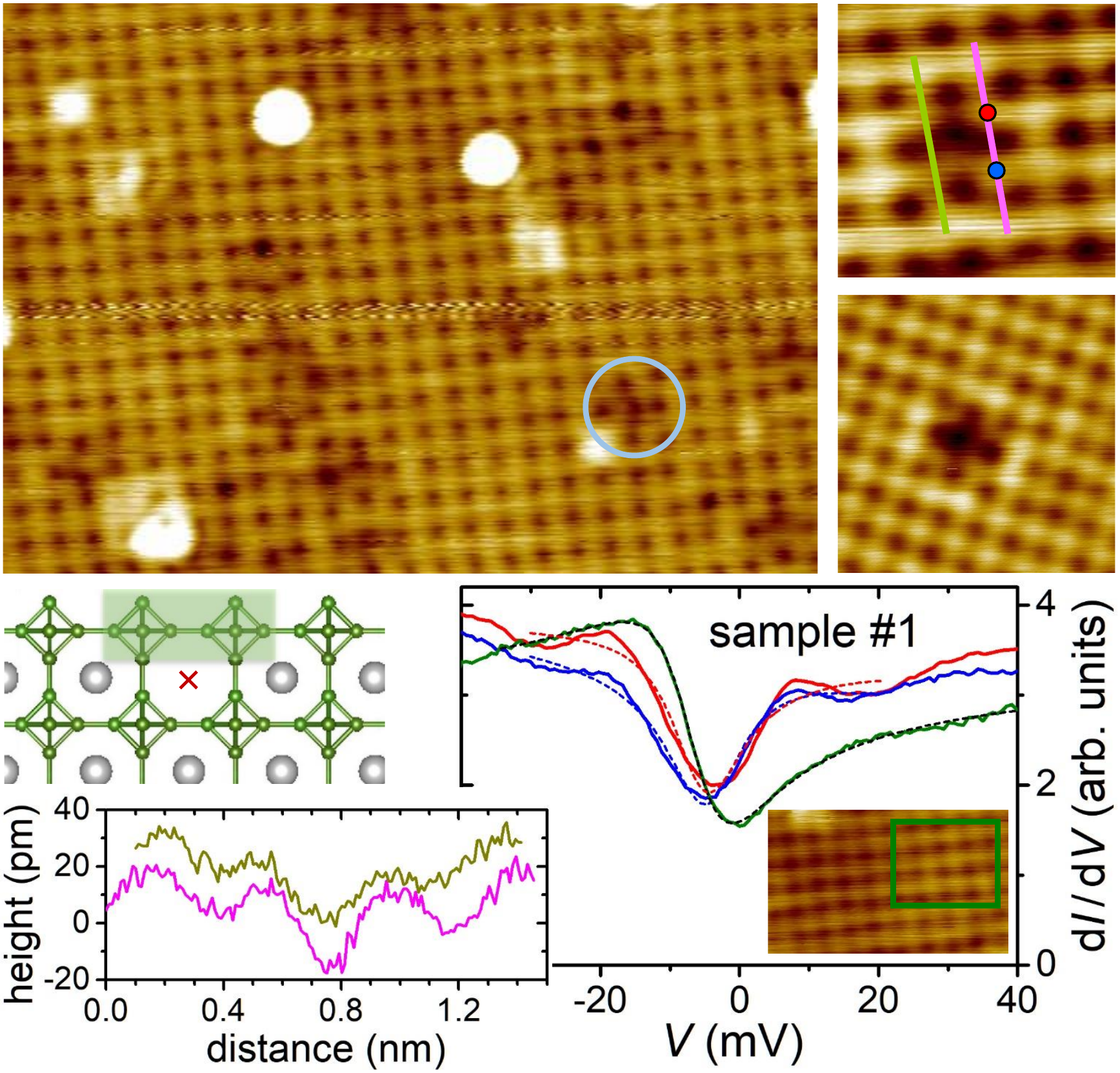}
\unitlength1cm \begin{picture}(-0.4,2)
\put(-8.6,7.8){\sffamily\bfseries\large \textcolor{white}{(a)}}
\put(-0.8,7.8){\sffamily\bfseries\large \textcolor{white}{(b)}}
\put(-0.8,5.6){\sffamily\bfseries\large \textcolor{white}{(f)}}
\put(-5.78,3.3){\sffamily\large (c)}
\put(-7.8,0.9){\sffamily\large (d)}
\put(-4.95,2.5){\sffamily\large (e)}
\end{picture}
\caption{(a) Topography (7 nm$\,\times\,$10 nm) showing several cross-like
defects on sample \#1 (one marked by blue circle, $V_{\rm b} = +0.07$~V,
$I_{\rm sp} =$ 0.3 nA). (b) Zoom into such a defect, area 2 nm$\,\times\,$2 nm
not included in (a). (c) Side view of possible defect formation by missing Sm
($\times$ mark). (d) Height profiles along the lines marked by similar colors
in (b). (e) Red and blue spectra taken at the respective points marked in (b)
within the defect. Green spectrum (5$\,\times\,$4 grid average) obtained in
the undisturbed area (green square) shown in the inset, 4 nm$\,\times\,$2.4
nm. Dashed lines are Fano fits. Set point: $V_{\rm b} = +0.07$~V,
$I_{\rm sp} =$ 0.3 nA. (f) Similar defect on pristine SmB$_6$, 3$\,\times \,$3
nm$^2$, $V_{\rm b} = +0.2$~V, $I_{\rm sp} =$ 0.6 nA, $T =$ 5.9 K.}
\label{cross}  \end{figure}
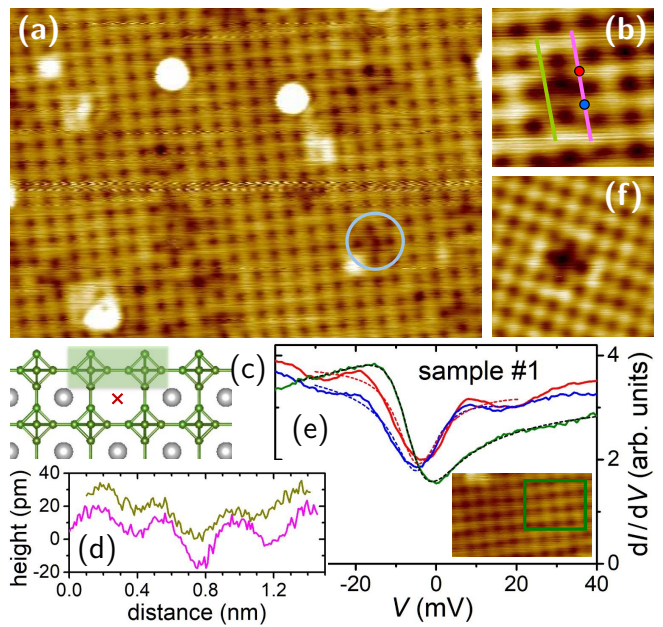
defect in Fig.\ \ref{cross}(a) is a cross-like dent, see blue circle. The
zoomed view of a different area, Fig.\ \ref{cross}(b), clearly reveals that
four atoms in a square arrangement of size $a^2$ are shifted lower into the
surface by about 10 -- 15 pm, cf. line scans in Fig.\ \ref{cross}(d). It is
important to note here that this type of defect is fairly regularly encountered
on surfaces of Sm-deficient samples. In contrast, on stoichiometric SmB$_6$,
for which we have reported topographies of 24 cleaves \cite{wir20}, we only
found one matching topography, presented in Fig.\ \ref{cross}(f). This
statistics makes a link between Sm-deficiency and the occurrence of these
defects likely. In particular, a missing Sm atom in a sub-surface layer, which
may easily be present in the Sm-deficient samples, may cause the four adjacent
B$_6$-octahedra within the top layer to slightly rearrange [in the side view,
Fig.\ \ref{cross}(c), the missing Sm is marked by $\times$, adjacent octahedra
in the top layer are shaded green]. Within the field of view of Fig.\
\ref{cross}(a) there are about 14 such defects. If they are all indeed due to
missing Sm-atoms, about 3.4\% of Sm would be absent in this particular layer,
which is around twice as much as statistically expected. However, as a cleave
certainly takes place at structurally weakened positions, such a deviation is
conceivable.

By comparison \cite{roe14,wir20}, the topographies presented in Fig.\
\ref{cross}(a) and (f) represent B-terminated surfaces, as schematically
depicted in Fig.\ \ref{cross}(c). In this case, defects (missing atoms) on
B$_{(6)}$ or B$_{(2)-(5)}$ sites should result in single or, possibly, double
dents. Moreover, the spectra obtained at the defect side, red and blue line in
Fig.\ \ref{cross}(e), as well as on a cleaner spot of the same sample (green
line and inset) support the assignment to a B-terminated surface. The possible
tunneling into Sm 4$f$ states as well as the conduction band results in a
co-tunneling phenomenon. In the simple Fano picture \cite{fan61,sch00}, the
tunneling conductance can be described by
\begin{equation}
\frac{{\rm d} I}{{\rm d} V} \propto \frac{(\epsilon + q)^2}{\epsilon^2 + 1}
\; , \qquad \epsilon = \frac{2(eV - E_0)}{\Gamma} \: , \label{fanf}
\end{equation}
where $\Gamma$ is the resonance width and $E_0$ the position in energy
relative to $E_{\rm F}$ may be influenced by the two tunneling channels.
Importantly, the asymmetry parameter $q$ depends on the ratio of tunneling
probabilities into the 4$f$ states vs. into the conduction band, and on the
particle-hole asymmetry of the conduction band \cite{fig10}. While a peak at
small negative bias voltage \cite{roe14,ruan14,jiao16,sun18,pir20,her20}
indicates tunneling into 4$f$ states, we only observe a small hump around
$-20$~mV. Fits to eq.\ \ref{fanf} (blue dashed line) yield $|q| \approx 0.18$
($|q| \approx 0.1$) for the red (blue) spectrum in Fig.\ \ref{cross}(e), i.e.
very little tunneling into the 4$f$ states. For comparison, $|q| \approx 0.7$
on the clean surface area shown in the inset of Fig.\ \ref{cross}(e), in good
agreement with \cite{roe14}. We speculate that the lower $q$-value in the
Sm-deficient sample is related to the missing Sm. In addition, the fit to the
spectrum on a clean side works nicely, while being considerably less reliable
at the defect. This may indicate a less-developed Kondo hybridization at the
defect site. Such a conjecture is supported by the resonance widths: fitting
the spectrum of the undisturbed area (green line and black dashed line) yields
$\Gamma \approx$ 15.1 meV in line with earlier results \cite{zha13,neu13,
roe14}, while it appears somewhat reduced within the defect (red dashed line:
$\sim$12.9 meV; blue: $\sim$13.4 meV). Hence, the hybridization gap is
reduced at these defect sites. It is also important to note that the trend
we observed for the position of the hump near $-20$ mV agrees well with
the report by Sun et al.\ \cite{sun18}: The peak moves to more negative
energies and gets less pronounced in height if less clean positions are
investigated. Also, there appears to be a small shift of the minimum in d$I
/$d$V$ from $E_{\rm F}$ towards slightly negative energies at surface areas
with impurities and/or adatoms, in line with data presented in \cite{sun18}.

\subsection{STM on sample \#2}
While the nominal Sm deficiency of samples \#2 is more pronounced compared to
\#1, the refined Sm occupancy is very close to unity, see Tab.\ \ref{tab1}.
In line with the latter and as exemplary shown in Fig.\ \ref{topo075},
atomically flat surface areas with only a surprisingly small number of defects
can be found on sample \#2, just as rarely as on pristine SmB$_6$. These
topographies are consistent with ($1\times 1$) surface terminations
encountered on stoichiometric SmB$_6$. While the off-stoichiometry of our
samples Sm$_{x}$B$_6$ certainly influences their cleaving process (in accord
with a faster polishing \cite{eo19}), it appears it does not prevent the
SmB$_6$ to form over reasonably large areas, similar to our observations on
sample \#1. However, as the cleave likely proceeds along crystalline defects
and may leave the pristine SmB$_6$ intact, we refrain from any statistics of
how much of the surface area might point toward an underlying little-disturbed
SmB$_6$ phase. Note that the very faint inhomogeneity (of a few lattice
constants in extend) below the atomic protrusions seen within the clean
\begin{figure}[t]
\includegraphics[width=0.4\textwidth]{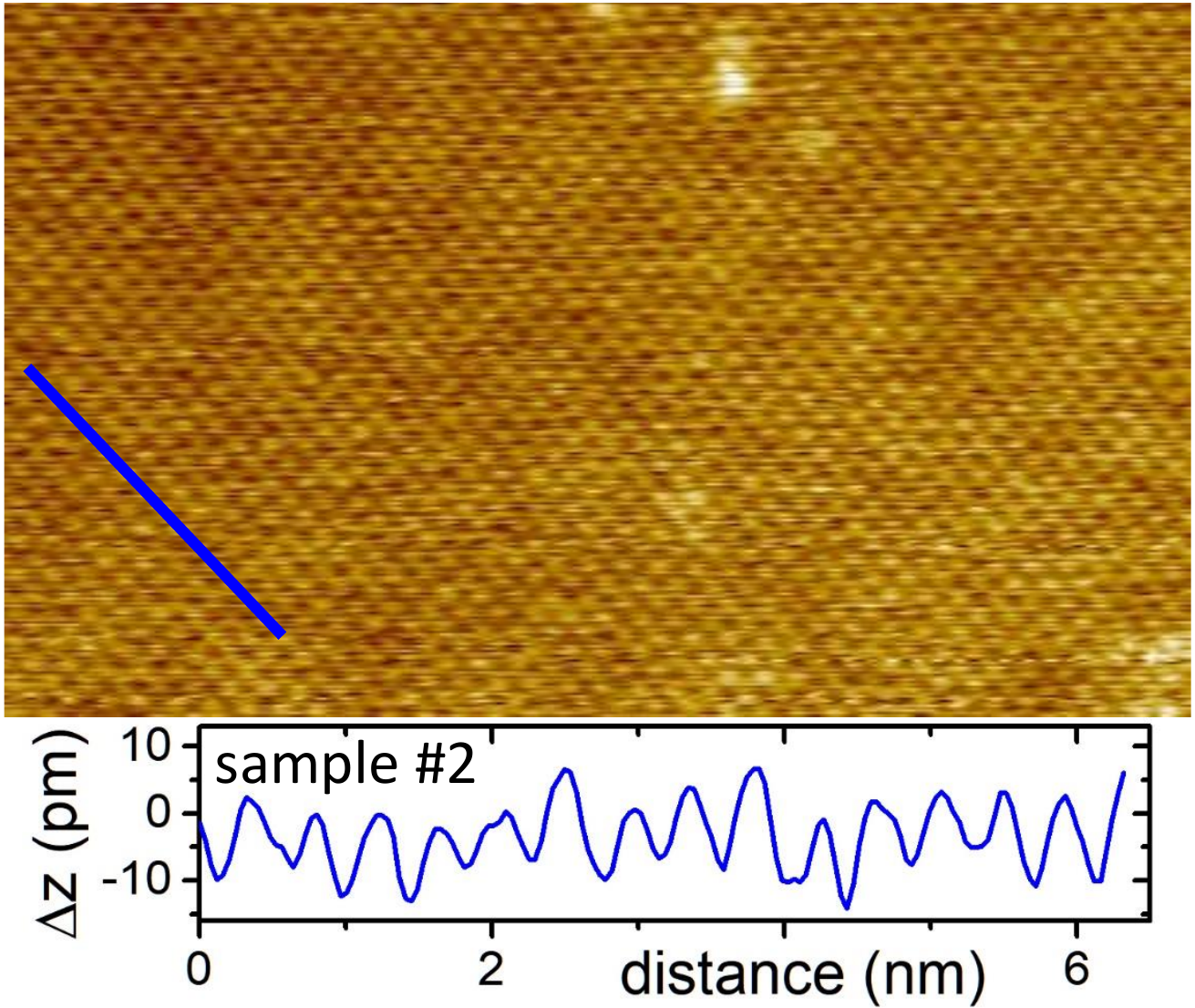}
\caption{Topography on sample \#2 within a relatively clean area of 20
nm$\,\times\,$12 nm. $V_{\rm b} = +0.07$~V, $I_{\rm sp} =$ 0.14 nA. The height
scan (averaged over thickness of the blue line) evinces atomic resolution.
This specific area does not show any sign of significant Sm-deficiency.} 
\label{topo075}  \end{figure}
($1\times 1$) surface of sample \#2 is also reminiscent to observations on
clean SmB$_6$ surfaces \cite{roe14,roe16,pir20}. So far, such inhomogeneities
were only encountered at temperatures around 5 K or above, but not below 2 K,
which may indicate a not completely formed conducting surface state at
$T \approx 5$ K.

To further scrutinize the above assumption of reasonably large SmB$_6$ regions
again a step edge is investigated. Atomic resolution within the terraces
indicate $(1 \times 1)$ terminations separated by a step of about 240 pm in
height, see Fig.\ \ref{spec075}(b). The spectra obtained within clean areas
of these two terraces are very similar, with minor differences on the negative
bias side. The Fano fits (dashed lines in Fig.\ \ref{spec075}(c)) yield
$\Gamma \approx$ 16.6 meV (15.7 meV) for the red (green) spectrum, in good
agreement with results on SmB$_6$ and clean surface areas of sample \#1,
indicating a well-developed hybridization gap. The $q$-values are small, $|q|
\approx 0.25$ and 0.34, with the smaller one on the upper terrace (red
spectrum) indicating suppressed tunneling into the Sm $f$-states. In addition,
the barrier heights are similar on both terraces, $\Phi \approx$ 4.8~eV and
4.3 eV, see Fig.\ \ref{spec075}(d). Taken together, we surmise that both
terraces of sample \#2 in Fig.\ \ref{spec075}(a) depict surfaces of the
SmB$_6$ phase, likely with B-termination. The upper (left) terrace would then
correspond to the B$_{(1)}$-surface encountered also on SmB$_6$, while the
lower one (right) may be a B$_{(6)}$-surface. In this case, the expected step
height is 0.248 nm, in good agreement with our measurements. Interestingly,
such an assignment could even account for the apparent lesser height of the
corrugations on the lower B$_{(6)}$-terrace compared to the upper
B$_{(1)}$-region, Fig.\ \ref{spec075}(b).

A B$_{(6)}$-surface has been considered unlikely based on surface energy
calculations \cite{sun18}. On the other hand, the cluster surfaces favored by
these calculations seem inconsistent with the majority of the topographies
observed here. It should be kept in mind, however, that the Sm-deficiency of
our samples most likely influences the samples' cleavage behavior; they cleave
more easily compared to pristine SmB$_6$ samples. This is further corroborated
by the observation of a new type of defect as well as a new surface
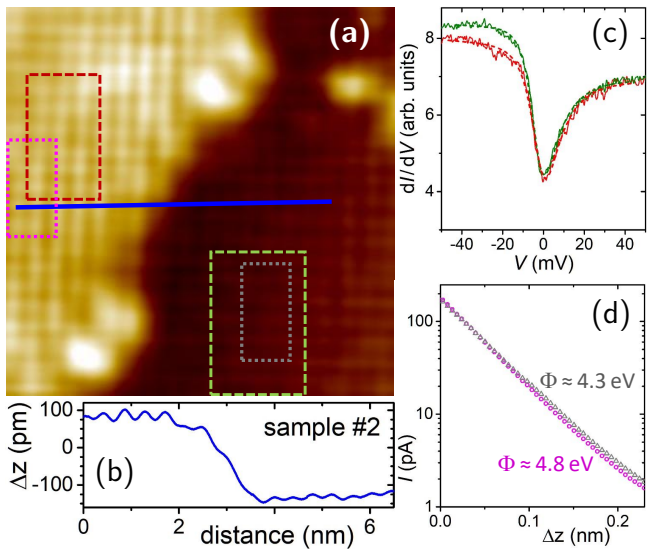
\begin{figure}[t]
\hspace*{-0.3cm}\includegraphics[width=0.48\textwidth]{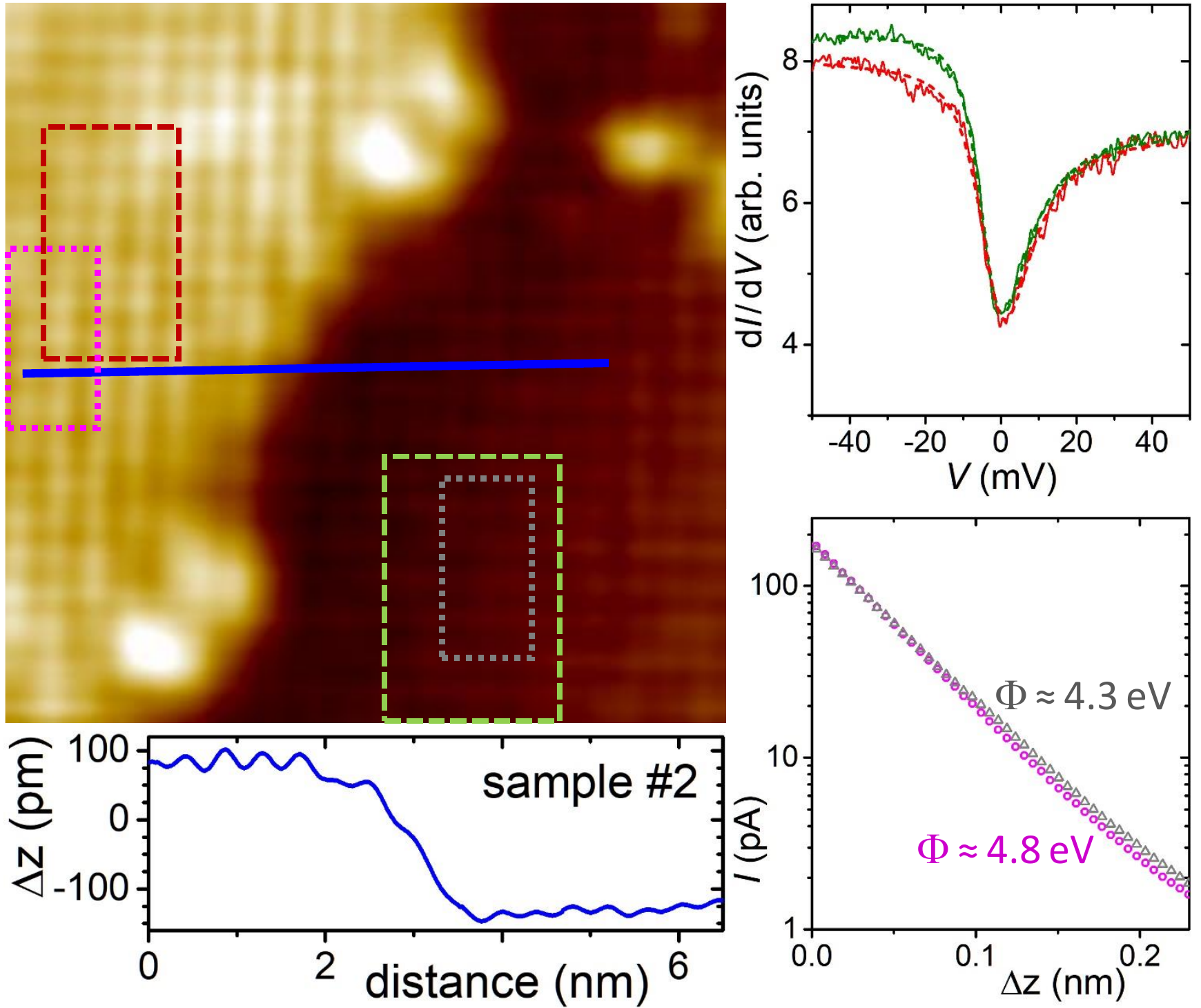}
\unitlength1cm \begin{picture}(-0.4,2)
\put(-4.4,6.75){\sffamily\bfseries\large \textcolor{white}{(a)}}
\put(-7.5,0.9){\sffamily\large (b)}
\put(-0.9,6.75){\sffamily\large (c)}
\put(-0.9,3.0){\sffamily\large (d)}
\end{picture}
\caption{(a) Step edge on sample \#2 within an area of 8 nm$\,\times\,$8 nm
($V_{\rm b} = +0.07$~V, $I_{\rm sp} =$ 0.14 nA). (b) Height scan along the
blue line marked in (a). (c) Averaged spectra taken within rectangles (dashed
lines) of corresponding colors in (a). Dashed lines are Fano fits. Set point:
$V_{\rm b} = +0.07$~V, $I_{\rm sp} =$ 0.14 nA. (d) $I(\Delta z)$-curves
obtained within areas marked by dotted lines in (a). 2 $\times$ 3 individual
curves at positions 1 nm apart were averaged ($V_{\rm b} = +0.07$~V).}
\label{spec075}   \end{figure}
reconstruction on surfaces of sample \#2, which were not encountered on
any of the more than 30 cleaved SmB$_6$ surfaces \cite{wir20}. Taken together,
it appears the surface energy is only one of the parameters determining the
cleaved surfaces, and may even be changed with respect to pure SmB$_6$.

\section{Discussion}
Interestingly, the vast majority of atomically resolved surface topographies
point to a seemingly undisturbed SmB$_6$ surface structure, as clearly shown
in Fig.\ \ref{topo075}. This finding is consistent with our XRD results.
Albeit the topographies presented in Figs.\ \ref{topo42} and \ref{cross}
highlight defects specific to Sm$_x$B$_6$ with $x<1$ (i.e.\ which are not or
very rarely encountered on stoichiometric SmB$_6$ surfaces), the underlying
topography away from defects appears highly similar to those found on
SmB$_6$ (see, e.g.\ Ref.~\onlinecite{wir20}). In fact, atomically resolved
surface areas were found somewhat more easily on Sm-deficient samples in
comparison to stoichiometric SmB$_6$, even though extensive search was
still required. Possibly related to this issue, a step edge as presented in
Fig.\ \ref{topo09} could so far only be found on one Sm-deficient sample of
batch \#1. Given the step height and the properties of the adjacent surfaces,
it is consistent with a line defect. We note here that, as STM topography
only depicts the two-dimensional sample surface, we cannot distinguish whether
these observed defects derive from grain boundaries or dislocations.
Despite extensive search, pristine SmB$_6$ (more than thirty cleaves) and
samples of batch \#2 did not reveal such a step edge.

Thermodynamically, the Sm:B solution has a large negative enthalpy of mixing
at the 1:6 composition. Therefore, for flux-grown samples an actual
composition of SmB$_6$ can be expected. Nonetheless, modified material
properties of flux-grown Sm$_x$B$_6$ have been reported for $x < 1$, e.g. in
Hall measurements \cite{sta19} and microhardness \cite{eo19}. Together with
the observations of line defects (Fig.\ \ref{topo09}) and a modified cleaving
behavior of the Sm-deficient samples (Sm$_x$B$_6$ samples with $x < 1$ require
much less force for cleaving compared to those with $x = 1$) we speculate that
the off-stoichiometry of Sm and B in the flux results in an increased
granularity of the samples while the SmB$_6$ stoichiometry is rather closely
preserved within the grains.

The majority of the surfaces discussed here are B-terminated. In this respect
it is worth noting that the different reports agree on their assignment of
the B-terminated surfaces \cite{roe14,yee13,ruan14,sun18,pir20,her20,wir20}.
Our observation of a cross-like defect on sample \#1, Fig.\ \ref{cross}
further confirms this assignment. In contrast, the Sm-terminated and the
$(2 \times 1)$ reconstruction are still under debate. So far, we could not
unambiguously identify a Sm-terminated surface on Sm-deficient samples.
However, we observed a new type of surface reconstruction in one instance
(therefore, it is not presented here), which we tentatively assigned
$c$($\sqrt{2} \times 3\sqrt{2}$)R45$^{\circ}$. Apparently, the SmB$_6$
structure is preserved locally, while an overall, crystallite-like structure
prevails due to the Sm-deficiency.

The observed barrier heights on our Sm-deficient samples are mostly around
5 eV, in good agreement with reports on pristine SmB$_6$ \cite{sun18} and even
EuB$_6$ \cite{wir20}. For a Sm-terminated surface, a very low work function
of order 2 eV is predicted \cite{sun18}. This should be kept in mind since
tunneling is limited to $V_{\rm b} \ll \Phi_{\rm s,t}$, i.e. the bias voltage
should not exceed a few tenths of one V.

It is interesting to note that, within clean areas of likely B-termination,
the resonance width is about 15 -- 16 meV, in good agreement with results on
pure SmB$_6$ samples \cite{roe14,sun18}. This value is somewhat reduced at
defects, see Fig.\ \ref{cross}(e). Spectra obtained at small spots between
B-clusters exhibited also a reduction in gap size \cite{sun18}. This
indicates that the electronic properties of Sm-deficient samples, in
particular the hybridization gap, are globally very similar to SmB$_6$, and
influenced only locally by defects or off-stoichiometry.

As discussed above, section \ref{sec-res}, transport measurements were
conducted on one bent sample \#1. The concomitantly increased $\rho(T)$
at room temperature might be explained, according to \cite{ste17}, by an
increased Sm valence with tensile strain, which enhances scattering in the
Sm 4$f^6 \leftrightarrow 4f^5 + 5d$ channel, and stronger hybridization
between $f$- and $d$-orbitals. STS taken beforehand on this surface could
nicely be fit by eq.\ (\ref{fanf}) with $|q|$-values as large as 0.83, which
is larger than any value we obtained on B-terminated surfaces so far
\cite{roe14,jiao16} (fits yielded 14.6 meV $\leq \Gamma \leq$ 16.7 meV).
However, at present we cannot directly correlate this observation with the
bending of the sample.

\section{Conclusion}
Macroscopic and microscopic (down to the atomic level) studies were combined
on the same Sm-deficient samples Sm$_x$B$_6$ with nominal deficiencies up to
25\%, i.e. $x =$ 0.75. Despite this high nominal Sm deficiency, the SmB$_6$
structure in the flux-grown samples remains strikingly stable such that
the hybridization gap and the low-temperature surface conductance appear not
significantly altered from pure SmB$_6$, an observation supporting a recently
reported protection of the SmB$_6$ transport gap against disorder \cite{eo19}.
The Sm vacancies do not disturb the {\em global}, macroscopic properties of
SmB$_6$, but rather induce crystallographic defects and locally reduce the
hybridization gap at these defects. The STM topographies are in good agreement
with the XRD results: on sample \#2, for which the refined composition is
closer to the 1:6 stoichiometry than for sample \#1 (see Tab.\ \ref{tab1}), we
were able to find large areas with a smaller number of defects (e.g.\ Fig.\
\ref{topo075}) compared to our topographies on sample \#1. We also note
that for Sm$_x$B$_6$ with $x < 1$, XRD did not reveal any other phase than
SmB$_6$.

\section*{Acknowledgments}
Work at Los Alamos National Laboratory was performed under the auspices of
the U.S. Department of Energy, Office of Basic Energy Sciences, Division of
Materials Science and Engineering. ZF acknowledges support from the LANL
G.\ T.\ Seaborg Institute. Funding by the German Research Foundation (DFG)
through grant WI 1324/5-1 is gratefully acknowledged.

\end{document}